\documentclass[aps,pra,preprint,superscriptaddress,]{revtex4-2}

\usepackage[utf8]{inputenc}

\usepackage{graphicx}
\usepackage{mathtools}
\usepackage{booktabs}
\usepackage{listings}

\begin{document}

\title{An electron-beam based Compton scattering x-ray source for probing high-energy-density physics}

\author{Hans G. Rinderknecht}
\email[]{hrin@lle.rochester.edu}
    \affiliation{Laboratory for Laser Energetics, University of Rochester, Rochester, NY 14623-1299}
    \author{G. Bruhaug}
    \affiliation{Laboratory for Laser Energetics, University of Rochester, Rochester, NY 14623-1299}

\author{V. Mu\textcommabelow{s}at} % don't need to include my middle name here, so just Vlad Musat
    \affiliation{Department of Physics, University of Oxford, Oxford OX1 3PU, UK}
\author{G. Gregori}
    \affiliation{Department of Physics, University of Oxford, Oxford OX1 3PU, UK}
\author{H. Poole}
   \affiliation{Department of Physics, University of Oxford, Oxford OX1 3PU, UK}
\author{G. W. Collins}
    \affiliation{Laboratory for Laser Energetics, University of Rochester, Rochester, NY 14623-1299}
    \affiliation{Departments of Mechanical Engineering and Physics and Astronomy, University of Rochester, Rochester New York 14627, USA}
    \affiliation{Department of Physics, University of Oxford, Oxford OX1 3PU, UK}
\date{\today, work in progress}

\begin{abstract}
	The physics basis for an electron-beam--based Compton scattering (ECOS) x-ray source is investigated for single-shot experiments at major high energy density facilities such as the Omega Laser Facility, National Ignition Facility, and Z pulsed power facility. %diffraction, Thomson scattering, and imaging experiments on the OMEGA laser.%
	A source of monoenergetic ($\delta\epsilon/\epsilon < 5\%$) 10- to 50-keV x-rays can be produced by scattering of a short-pulse optical laser by a 23- to 53-MeV electron beam and collimating the scattered photons.
	The number and spectrum of scattered photons is calculated as a function of electron packet charge, electron and laser pulse duration, laser intensity, and collision geometry.
	A source delivering greater than 10$^{10}$ photons in a 1-mm-radius spot at the OMEGA target chamber center and 100-ps time resolution is plausible with the available electron gun and laser technology.
	Design requirements for diffraction, inelastic scattering and imaging experiments as well as opportunities for improved performance are discussed.
\end{abstract}

\maketitle

\section{Introduction}
\label{sec:intro}
Understanding the structure and dynamics of dense matter at moderate temperature is at the cutting edge of high-pressure physics and has important applications for research in planetary interiors, inertial fusion, and advanced materials manufacturing \cite{Remington2005, Kraus2022, Leung2018, Abu2022}. 
Laboratory experiments using high-power laser facilities are now able to access extreme material conditions with pressures exceeding 10 to 100 Mbar (1-10 TeraPascal) and with temperatures ranging from a low ($\sim$0.1 eV) quantum dominated regime \cite{Nature:Smith:2014} to a moderately kinetic warm-dense-matter regime ($\sim$10 eV), where the thermal, Coulomb, and Fermi energies of the conduction electrons are all comparable \cite{Kritcher2020}.  
Recent discoveries in the low-temperature quantum regime reveal that dense matter in these conditions can behave quite differently than expected from the longstanding Thomas-Fermi statistical model, giving way to structural and electronic complexity and coherence.  While there are a growing number of predictions for such behavior for elements and compounds \cite{Hilleke2022,Pickard2011,Zurek2011}, few data exist \cite{Polsin2022,Gorman2022} because of the limiting quality of x-ray sources at the major compression facilities. 

The warm dense matter regime represents a significant theoretical and computational challenge as traditional condensed matter techniques are only applicable to cold (i.e., with temperatures well below the Fermi level) systems, while classical plasma expansion approaches are also inapplicable since the matter remains strongly correlated. The main difficulties in modeling warm dense matter states are finite-temperature electron degeneracy and strong interparticle correlations, requiring a full quantum mechanical treatment of the free electrons while retaining a many-body description of the ion motion together with exact calculations of the bound and valence orbitals.
All the computational techniques employed so far (see, e.g., Refs. \cite{Ichimaru1982,Jones2015,White2013,Dornheim2018,Larder2019}) have used approximations which are largely untested as experimental benchmarks are sparse \cite{Glenzer2009}.  
Moreover, with the recent advances in machine-learning methods, it has become clear that progress in this field relies on data-driven approaches that have the potential to explore much wider parameter space and exploit new relationships that have so far remained hidden in our current physical models \cite{Hatfield2021}. 

While the availability of a large number of accurate experimental data sets is therefore important for progress, there is still no facility worldwide where these explorations can be performed. 
Free-electron laser (FEL) facilities have exquisite diagnostic capabilities thanks to the availability of collimated, high-brightness short pulses of x rays, but they lack the capability to produce extreme matter conditions, except those in the lower-pressure and -temperature regimes \cite{Fletcher2015}.
On the other hand, high energy density facilities, such as Omega, the National Ignition Facility, or Z, excel in accessing a wide range of conditions, but lack advanced x-ray probing capabilities --- mostly limited to noncollimated, incoherent atomic fluorescence sources produced by thermal ionization or fast electron heating.
While, ideally, the combination of FEL's and high-energy-density compression capabilities into a single multipurpose facility would be able to address the above needs, such a facility will require a substantial capital investment and it is unlikely it would become available in the near term.

This paper explores a different approach to making a single shot, collimated, narrow bandwidth x-ray source available for diagnosing experiments at the Omega Laser Facility, at a significantly reduced cost. 
Here we propose to use conventional linac technology for the generation of a 23- to 53-MeV electron beam and then employ inverse Compton scattering from an optical high-intensity laser for the generation of a 10- to 50-keV x-ray impulse containing at least $10^{10}$ photons with less than 5\% bandwidth and duration of less than 100~ps.
This technology can be readily used for warm dense matter diagnostics, and will have the potential to open a new frontier for discovery science in high-energy-density (HED) physics.

The manuscript is organized as follows:
Section~\ref{sec:phys} introduces the relevant physics underlying Compton scattering as an x-ray source.
Section~\ref{sec:source} discusses in more detail the electron beam and laser properties required for the source, as well as constraints on the beam-laser interaction point, and presents an estimate for the performance of such a source based on existing linac and laser technologies.
Section~\ref{sec:sims} presents simulations that test and confirm the analytical estimate of the proposed Compton source performance.
Section~\ref{sec:implementation} explores the requirements for integrating such a source with the existing Omega Laser Facility target areas.
Finally, Section~\ref{sec:apps} investigates the applicability of the proposed source for a variety of HED diagnostic techniques.

\section{Physics Basis}
\label{sec:phys}
Compton scattering is the canonical electrodynamic phenomenon of a charged particle scattering a high-energy photon.  
In the case of a relativistic electron beam with Lorentz factor $\gamma = \left(1-\beta^2\right)^{-1/2} \equiv (1 + E_{\textrm{e}}/mc^2)$ interacting with a laser pulse, the apparent frequency of the photons in the reference frame of the electrons is increased by a factor $(1+\beta)\gamma$. 
Assuming the electron's momentum is not significantly changed and the photon is forward scattered, returning to the laboratory frame applies this multiplier again, for a total increase in frequency and energy of approximately $4\gamma^2$.
This quadratic scaling allows scattered optical photons ($\epsilon_{\textrm{i}}\sim1~$eV) to reach the x-ray regime ($>$1~keV) by scattering from an electron beam with $\gamma \gtrsim 16$ ($E_{\textrm{e}} \gtrsim 8$~MeV).

Accounting for relativistic electron orbits in a counter-propagating intense laser field, the scattered photon has a wavelength $\omega_{\textrm{f}}$ that depends on the initial laser wavelength $\omega_{\textrm{i}}$ and other terms as \cite{FIP:Alejo:2019}:

\begin{equation}
	\omega_{\textrm{f}} \approx \frac{2 \gamma^2 \omega_{\textrm{i}} \left(1 + \cos{\phi}\right) N_{\textrm{p}} }{1 + \gamma^2\theta^2 + \frac{a_0^2}{2}+2\frac{\chi N_{\textrm{p}}}{a_0}}.
	\label{eq:frequency}
\end{equation}
\noindent Here, $\phi$ is the incident angle of the laser, $a_0 = eE/\omega_{\textrm{i}}mc\approx0.86\sqrt{I_{18}\lambda_{{\mu}\textrm{m}}^2}$ is the normalized vector potential of the incident laser with intensity $I_{18}$ in units of 10$^{18}$~W/cm$^2$ and wavelength $\lambda_{\mu\textrm{m}}$ in microns; $N_{\textrm{p}}\approx\textrm{max}(1,a_0^3)$ represents the number of photons scattered per event; $\theta$ is the angle of the scattered photon relative to the electron-beam direction, and $\chi\approx\gamma a_0/a_{\textrm{c}}$ represents the laser electric-field strength in the electron rest frame normalized to the critical field amplitude $a_{\textrm{c}} \approx 4.1\times10^5 \lambda_{{\mu}\textrm{m}}$.
(For conditions discussed in this work, the last term in the denominator will be negligible.)

\begin{figure}
	\includegraphics[width=\textwidth]{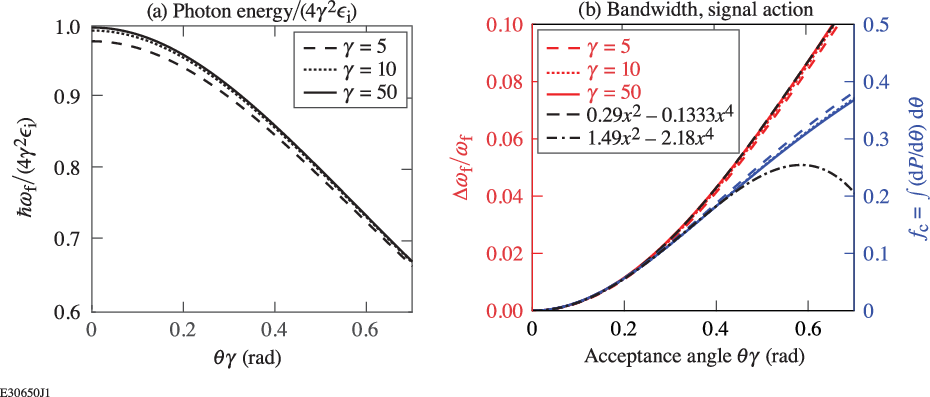} %{E_bw_P_vs_theta.png}
	\caption{(a) Normalized photon energy gain as a function of the product of detection angle $\theta$ and electron-beam Lorentz factor $\gamma$; (b) fractional bandwidth (red) and cumulative signal fraction (blue) for collimation acceptance angle $\theta\gamma$.  Calculations assume head-on scattering ($\phi = 0$).\label{fig:analyticmodel}}
\end{figure}

Plots of the scattered photon energy in the limit of head-on scattering $(\phi = 0)$ are shown in Fig.~\ref{fig:analyticmodel}(a).
Although the scattered photons are monoenergetic at any single detection angle, collecting photons scattered into a range of angles will produce a broad-band source.
The bandwidth is then a function of the collection solid angle that scales as $(\Delta \omega_{\textrm{f}}/\omega_{f}) \approx 0.29(\theta\gamma)^2 - 0.13(\theta\gamma)^4$ in the range $\theta\gamma<0.7$, as shown in Fig.~\ref{fig:analyticmodel}(b).
Integrating over the differential cross section for scattering and transforming to the laboratory reference frame, the fraction of photons collected scales with collection solid angle as $f_{\textrm{C}} \approx 1.49(\theta\gamma)^2~-~2.18(\theta\gamma)^4$ in the range $\theta\gamma<0.4$.  
(Details of this calculation are given in Appendix~\ref{app:derivation:xs}.)
The efficiency of the source is limited by the required bandwidth: to achieve 1\% (2\%) intrinsic bandwidth requires collimation to $\theta\gamma\leq0.19$ (0.27) rad, which in turn includes only 5\% (10\%) of the scattered photons.
If the photon energy is tuned using the electron-beam energy, the collimation will need to be adjusted to maintain optimal collection angle as a function of $\gamma$.
Other sources of spectral broadening include \cite{SciRep:Kramer:2018}:

\begin{equation}
	\frac{\Delta\omega_{\textrm{f}}}{\omega_{\textrm{f}}} \approx \sqrt{\left(\frac{\Delta \omega_{\textrm{i}}}{\omega_{\textrm{i}}}\right)^2 + \left(\frac{2\Delta \gamma}{\gamma}\right)^2 + \left(\frac{0.88 a_{0,\textrm{eff}}^2}{2+a_{0,\textrm{eff}}^2}\right)^2 + \left(\frac{1.05\left(\gamma\sigma_{\theta,\textrm{eff}}\right)^2}{1 + \left(\gamma\sigma_{\theta,\textrm{eff}}\right)^2}\right)^2}
	\label{eq:bandwidth}.
\end{equation}
Here, $a_{0,\textrm{eff}}$ is the effective normalized vector potential weighted by the local number of photons in the laser pulse, and $\sigma_{\theta,\textrm{eff}}$ is the electron-beam divergence weighted over the laser pulse.
The x-ray bandwidth scaling with laser and electron bandwidth follows directly from the numerator of Eq.~(\ref{eq:frequency}).
The quadratic scaling with laser intensity $a_0$ arises from a reduction in the instantaneous energy of the electron beam within the laser packet due to the ponderomotive force.
(This form assumes a Gaussian laser packet.)
To maintain a scattered photon bandwidth of 1\%, Eq.~(\ref{eq:bandwidth}) suggests the intensity must be limited to $a_{0,\textrm{eff}} < 0.15$.  
With control of the laser temporal and spectral properties this effect may be limited, allowing monoenergetic scattering with more intense beams \cite{PRE:Ramsey:2022}.  
%However, intensity approaching $a_0 = 1$ will distort the electron beam propagation and introduce bandwidth in the scattered photon spectrum.
% 
%For interactions of an electron beam and laser with finite bandwidth and Gaussian temporal profile, the bandwidth of the emitted photons is approximately:

The number of photons scattered per laser cycle is given by \cite{FIP:Alejo:2019,RMP:Corde:2013}:
\begin{equation}
	\frac{N_x}{\nu_{\textrm{i}}^{-1}} \sim \left\{
	\begin{array}{@{}ll} 
		1.53\times10^{-2} a_0^2, & a_0 < 1 \\  
		3.31\times10^{-2} a_0, & a_0 \gg 1
	\end{array}
	\right. .
	\label{eq:N_scatter}
\end{equation}
\noindent The number of photons scattered by a relativistic electron charge packet interacting with a laser pulse can be estimated as the product of Eq.~(\ref{eq:N_scatter}) with the number of electrons in the packet ($N_{\textrm{e}}$) and the number of laser cycles ($N_{\tau}$).
The number of laser cycles observed by the electrons may depend on the temporal and spatial properties of the focused laser pulse.
Assuming a diffraction-limited focal spot, if the Rayleigh length $z_{\textrm{R}}$ is long compared to the pulse duration $\tau_{\textrm{L}}$ (that is, $z_{\textrm{R}} \approx 4 f_\#^2\lambda/\pi \gg \tau_{\textrm{L}} c$, for $f_\#$ the f-number of the focusing optic), then the temporal profile will limit the interaction, and the number of laser cycles will be $N_\tau = \tau_{\textrm{L}} \omega_{\textrm{i}} / 2\pi$. %and the number of scattering events will be:
%\begin{equation}
%	N_{x,tot,\tau} = 2.3\times10^{10} f_{\textrm{C}} \left(\frac{Q_e}{\textrm{1~nC}}\right) \left(\frac{\tau_{\textrm{L}}}{\textrm{1~ps}}\right) \left(\frac{\hbar \omega_{\textrm{i}}}{\textrm{1~eV}}\right) 	
%	\left\{
%	\begin{array}{@{}ll} 
%	    a_0^2, & a_0 < 1 \\  
%		2.2 a_0, & a_0 \gg 1
%	\end{array}
%	\right.
%	\label{eq:N_source}
%\end{equation}
Otherwise, the geometry of the interaction will limit the number of laser cycles to $N_\tau \approx 2z_{\textrm{R}}/\lambda = 8 f_\#^2/\pi$, or $0.75 f_\#/\phi$, whichever is smaller.
(The derivation of the geometric terms is described in Appendix~\ref{app:derivation:geometry}.)
In these cases, assuming the laser is in the regime $a_0 < 1$, the number of scattering events is approximately
%\begin{equation}
%	N_{x,tot,z} = 1.5\times10^{10} f_{\textrm{C}} \left(\frac{Q_e}{\textrm{1~nC}}\right) \left(\frac{f_\#}{10}\right)^2 \min\left[\frac{2}{\pi f_\# \phi}, 1\right]
%	\left\{
%	\begin{array}{@{}ll} 
%		a_0^2, & a_0 < 1 \\  
%		2.2 a_0, & a_0 \gg 1
%	\end{array}
%	\right.
%	\label{eq:N_source_z}
%\end{equation}
\begin{align}
	N_{x,\textrm{tot}} &= f_{\textrm{C}} N_e \left(\frac{N_x}{\nu_{\textrm{i}}^{-1}}\right) N_\tau \nonumber \\
	 &\approx 10^9 \left(\frac{\theta\gamma}{0.27}\right)^2 \left(\frac{Q}{1~\textrm{nC}}\right) a_0^2 \left\{\begin{array}{@{}ll}
		2.31 \left({\displaystyle\frac{\hbar \omega_{\textrm{i}}}{1~\textrm{eV}}}\right) 	\left({\displaystyle\frac{\tau_{\textrm{L}}}{1~\textrm{ps}}}\right), & \tau_{\textrm{L}} c \ll z_{\textrm{R}} \\
		2.43 \min\left[1, {\displaystyle\frac{0.294}{f_\# \phi}}\right] \left({\displaystyle\frac{f_\#}{10}}\right)^2, & \tau_{\textrm{L}} c \gg z_{\textrm{R}} \\
	\end{array}
	\right. .
	\label{eq:N_source}
\end{align}
\noindent Typically, the second condition will hold since $z_{\textrm{R}}/c = 0.42$~ps for $f_\# = 10$ and a 1-$\mu$m laser wavelength.
The number of scattered photons is maximized with high charge ($Q = eN_{\textrm{e}}$), high intensity, and longer focal lengths.
%Some of these terms are limited by experimental considerations.

The use of a flying-focus laser may improve the performance by decoupling the length of the scattering volume ($L$) from the radius of the focal spot ($\sigma_L$) \cite{NP:Froula:2018}.
This makes more efficient use of laser energy.
With a flying-focus pulse, the number of cycles is simply $L/\lambda$ and the number of scattering events is
\begin{equation}
	N_{x,\textrm{FF}} = 7.7\times10^{9} \left(\frac{\theta\gamma}{0.27}\right)^2 \left(\frac{Q}{\textrm{1~nC}}\right) \left(\frac{L}{1~\textrm{mm}}\right) \left(\frac{\hbar \omega_{\textrm{i}}}{\textrm{1~eV}}\right)
	\left\{
	\begin{array}{@{}ll} 
		a_0^2, & a_0 < 1 \\  
		2.2 a_0, & a_0 \gg 1
	\end{array}
	\right. .
	\label{eq:N_source_ff}
\end{equation}
Moreover, the bandwidth dependence on intensity ($a_0^2$) is produced by the gradient in intensity, and assumes a Gaussian pulse.
A flying focus can produce a roughly flat intensity equal to the peak intensity that travels with the electron packet over a long distance.
This should reduce the $a_0$-dependent bandwidth term by the ratio of the rise and fall region divided by the length: $2a_0/\max(\nabla a_0) L$.
However, the size of the electron packet that fits inside the co-moving intense region will be limited by the Rayleigh length of the laser.
This relationship is discussed in Sec.~\ref{sec:intersection}.

For Eq.~(\ref{eq:N_source}) and (\ref{eq:N_source_ff}), the useful fraction of scattered photons is limited by bandwidth considerations to roughly $f_{\textrm{C}} \lesssim 0.1$ [see Fig.~\ref{fig:analyticmodel}(b)].
Because the scattered photons travel at approximately the same speed as the electrons, the temporal resolution of the source will be set by the duration of the electron packet.
The charge available in a photoinjector electron gun is limited by space charge and scales with the duration of the packet ($\tau_{\textrm{e}} = w/c$, for packet width $w$). % with a correspondence of roughly 1~pC per ps.  
An optimal design would then have laser and electron pulse durations equal at approximately the desired temporal resolution to maximize both the bunch charge and the number of laser cycles.

%In the non-relativistic regime ($a_0 < 1$), the strongest dependence on photon number is the laser intensity.
In the following section we will consider available electron gun and laser technology to assess the potential for a single-shot source capable of producing high x-ray fluence (above 10$^{10}$) while maintaining low bandwidth (below 5\%).

\section{Source Properties}
\label{sec:source}
\subsection{Electron Photoinjectors}
\label{sec:sub:electrons}
Electron photoinjectors are a commercial technology enabling ultrafast MeV-scale electron bunches \cite{vendor:Radiabeam}.
In these systems, a UV laser (typically with $\mu\textrm{J}$ energy) irradiates a photocathode to produce electrons, which are then accelerated using a small radio-frequency (rf) waveguide to several MeV.  
The photoinjector and first acceleration stage (5~MeV) of a commercial system is typically less than 1~m in length.
Further accelerating sections can be introduced to reach higher energies, with typical acceleration gradients of
%G.B. edit
20 MV/m, but gradients as high as 100~MV/m are available.
%G.B. edit, should we cite Radiabeam here or can this be taken as common knowledge?
Magnetic optics may be used to improve and control beam quality, such as pulse compression or focusing.
Pulse temporal compression can be achieved using chicanes (a sequence of dipole magnets) or rf cavities, if desired.  
The primary parameters of interest for this study are the total packet charge, which directly affects the number of scattered photons [Eq.~(\ref{eq:N_source})], and the beam emittance, which affects the bandwidth and spatial resolution.

%% HGR add discussion of solenoid focusing?

%Is it worth citing the IAC 5 nC in 60 ps accelerator? The emittance is higher then we want, but not insanely so. https://accelconf.web.cern.ch/ipac2012/papers/tuppd079.pdf

\subsubsection{Packet charge and width}
Several examples exist in the literature of sources producing tens of nC of electrons in tens of ps bunches.
The A$\emptyset$ Photoinjector at Fermilab produces 16-MeV, 10- to 14-nC electron bunches with uncompressed length $< 40$~ps \cite{ILC:Fitch:2000}.
A subsequent compression by a nine-cell cavity rf stage can reduce the pulse length to under 10~ps.
The ELSA facility at CEA includes a photoinjector source that can produce 14-nC bunches in 90-ps lengths using a 1.2-$\mu\textrm{J}$, 60-ps laser pulse \cite{EPAC:Marmouget:2002}.
These bunches are accelerated to 16~MeV in the first stage of the linac ($\sim7$~m).
From the scaling in Eq.~(\ref{eq:N_source}), bunch charges on the 10-nC, 100-ps level are relevant to producing Compton x-ray sources with of the order of 10$^{10}$ photons.

For the purposes of HED physics experiments, the dynamics of interest are often on the scale of nanoseconds, which is the primary motivation for developing a bright single-shot electron beam source.
However, a multi-bunch photocathode rf gun system has been demonstrated \cite{NIM:Hirano:2006}, which may be useful for recording multi-frame `movies' of HED systems.
A 357-MHz (2.8-ns separation) pulse train of 266-nm UV ($4\omega$) laser light with 5~$\mu\textrm{J}$ per pulse was used to irradiate the photocathode.
The accelerating rf with a driving frequency of 2.856~GHz, or 8$\times$ the laser pulse frequency, was driven with up to 17-MW input power from a pulsed klystron.
Total charge up to 3.5~nC per packet was observed, accelerated to 5~MeV with 1\% momentum spread between packets.
This work suggests that such a system may provide the basis for a multipulse x-ray source with ns-scale pulse separation.
Since the resulting x-ray pulses would be colinear, a single line-of-sight time-resolving camera would be required to differentiate between signal pulses \cite{RSI:Theobald:2018}.
%Continuous Electron Beam Accelerator Facility (CEBAF) 'fills every bucket'

%Packet self force: a packet will fall apart on a timescale: 
%$F = m dV/dt = eE = e^2/(4 \pi \epsilon_0 R^2)$.
%Assuming the packet starts with radius $R_0$ and is 'disassembled' at $2 R_0$, we have:
%$ \int R^2 d^2R  = \int dt e^2/4 \pi \epsilon_0 m$

\subsubsection{Emittance}
Divergence of the electron beam $\sigma_\theta$ produces broadening in the spectrum proportionally to $\gamma\sigma_\theta$ [Eq.~(\ref{eq:bandwidth})].
Here, the divergence is defined as the rms average of the incident electron angle relative to the beam axis.  
The quality of an electron beam is usually characterized by the emittance $\varepsilon = \beta\gamma\sqrt{\langle x^2\rangle\langle x'^2\rangle - \langle x x'\rangle^2} \approx \beta\gamma\sigma_x\sigma_\theta$ in terms of the rms beam size $\sigma_x$ and divergence $\sigma_\theta$.
The measured emittance of the sources discussed above are 20$\pi$~mm~mrad (Fermilab A$\emptyset$) and $<$4~mm~mrad (CEA ELSA), respectively \cite{ILC:Fitch:2000,EPAC:Marmouget:2002}.
To limit spectral broadening due to beam emittance below 1\% (2\%) requires $\gamma\sigma_{\theta,\textrm{eff}}\lesssim0.1$~(0.14)~rad, respectively.
With an ELSA-quality beam, this level of divergence could be attained with a beam spot size $\sigma_x = \varepsilon / \beta\gamma\sigma_\theta \approx 40~\mu$m (29~$\mu$m) at focus.  
The smaller value between this and the laser focal spot size will define the x-ray source size, which in turn defines the resolution for imaging applications.

Lower emittance has been achieved in some systems with reduced bunch charges.  
The ELSA photoinjector achieved values as low as 1~mm~mrad at $Q$ = 1~nC, which was close to the thermal emittance of the cathode \cite{EPAC:Marmouget:2002}.
The BriXS Ultra High Flux inverse Compton source reports packets of 100 to 200~pC in 1.3 to 4.0-ps bunches with nominal normalized emittance in the range 0.6 to 1.5~mm~mrad \cite{Inst:Drebot:2019}.
A survey of the present literature suggested that comparably low-emittance bunches are limited to roughly the range 50- to 200-pC per ps pulse duration \cite{PhD:Tikhoplav:2006}.
In interactions with a flying focus, where shorter pulses are required, we will assume values of $\varepsilon = 1$~mm~mrad and $Q = (100~\textrm{pC/ps})\tau$ may be reasonably expected.

\subsubsection{Bandwidth}
Radio-frequency acceleration is self-correcting for electron energy dispersion and in general achieves very small bandwidth variations.
For example, the CEA ELSA accelerator described above produces 0.1\% rms energy dispersion \cite{EPAC:Marmouget:2002}.
Limiting the electron energy spread to less than 1\% should not be challenging, and the bandwidth of the resulting x-ray source should not be dominated by the ($\Delta\gamma/\gamma)$ term in Eq.~\ref{eq:bandwidth}.

\subsection{Laser Sources}
In the case of laser pulses with Gaussian temporal history, monoenergetic scattered x rays ($\Delta \omega_{\textrm{f}} / \omega_{\textrm{f}} \leq 1\%$) require a limit on the normalized vector potential of $a_0 < 0.15$, which limits the intensity as $I < 3\times10^{16}$~W/cm$^2 \left(\lambda_{\mu\textrm{m}}\right)^{-2}$.
This level of intensity in a 100-ps pulse duration with a focal spot of 40-$\mu$m radius would require 150~J of 1053-nm laser light with peak power of 1.5~TW.
Use of a higher-frequency laser at the same $a_0$ and $\tau_{\textrm{L}}$  would linearly increase the scattered photon frequency [Eq.~(\ref{eq:frequency})].
This would, however, require an increase in the laser intensity proportional to $\omega_{\textrm{i}}^2$.

A flying-focus pulse would make more-efficient use of the laser energy.  
To achieve an intensity of 3$\times10^{16}$~W/cm$^2$ in a focal spot with 40-$\mu$m radius and length of 1~mm would require roughly 5~J in a flying-focus configuration.  
However, the intense region would be a smaller region comoving with the electron packet.
%Because of this, an electron packet with 10 ps duration (3~mm in length) would mostly not overlap with the light, limiting the electron charge involved in the interaction.
The electron packet width would then be limited to the Rayleigh length of the focusing optic (in the example above, 1~mm/$c$ = 3.3~ps), which also limits the available charge.
This may still have an advantage if the lower-charge electron beam has improved emittance and interacts on average with a higher intensity laser packet.

\subsection{Beam Laser Interaction}
\label{sec:intersection}
\begin{figure}
	\includegraphics[width=\textwidth]{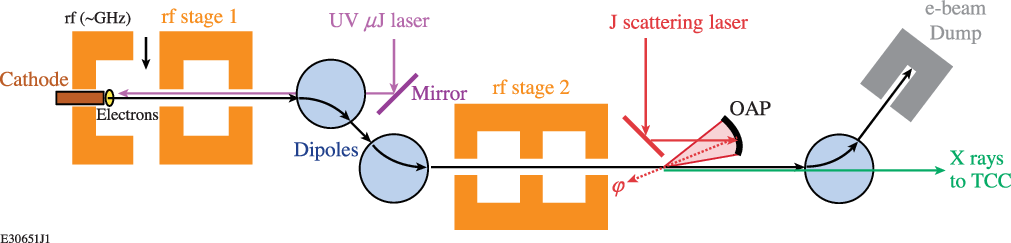} %{Fig1_layout.pdf} %, trim={60, 350, 60, 350}{2_v230209_Rinderknecht_2_9.pdf} %
	\caption{Schematic of an electron-beam Compton scattering x-ray source for OMEGA.\label{fig:cartoon}}
\end{figure}

A schematic diagram of an electron-beam based Compton x-ray source is shown in Fig.~\ref{fig:cartoon}.
The primary engineering challenge of the source is co-timing and co-aligning the electron beam and the scattering laser pulse.
At a minimum, the two beams must be co-timed better than the longer of the two pulse durations, and co-aligned better than the larger of the packet waist and the focal spot size.
Assuming the system is designed to achieve 100-ps temporal resolution, a timing jitter of the order of 10~ps will be required, equivalent to 3~mm of laser path.  
To robustly achieve this level of co-timing, a single laser front end may be used to seed both the $\mu$J UV laser that irradiates the photocathode and the scattering laser.

The creation of high-charge beams is more likely to be a limiting factor than laser intensity.
The most robust design will therefore feature a laser pulse that is both longer and spatially larger than the electron packet, such that the spatial resolution of the system is set by the electron packet size and the temporal resolution by the transit time.
Note that the geometric terms in Eq.~(\ref{eq:N_source}) strongly encourage an on-axis scattering geometry.  
If the laser is coaxial with the electron beam ($\phi = 0$), the geometric term for spatially limited scattering increases as $f_\#^2$, encouraging long scattering distances.  
But if the offset from on-axis scattering exceeds $\phi > 2/\pi f_\# (=3.6^\circ$~for $f_\# = 10)$, the intersection volume grows only linearly with $f_\#$. 
In the case of a flying focus [Eq.~(\ref{eq:N_source_ff})], the standard optics used for spatiotemporal control require on-axis scattering for the intense region to co-move with the electron packet.
For these reasons, an on-axis scattering geometry is highly beneficial for both cases.

%$t \approx \sqrt{\tau_{\textrm{e}}^2 + (w_L/c\sin\theta)^2}$ for beam waist $w_L$.
%If the laser is coaxial with the electron beam ($\theta = 0$), the latter term would be replaced by $\tau_{\textrm{L}}$ or the Rayleigh length divided by the speed of light, whichever is smaller.
%If a 1~$\mu$m scattering laser is focused with an f/20 optic to a diffraction-limited focal spot, such that the beam waist is 20~$\mu$m and the Rayleigh length 400~$\mu$m, and the pulse duration is much longer than the electron pulse duration, then the laser can be approximated as a high-field region that the electron packet transits in a time 

\subsection{Design calculation}
\label{sec:designcalcs}
On the basis of the above considerations, the performance of three plausible ECOS designs are detailed in Table~\ref{table:specs}.
In the case of a standard Gaussian laser focus, an ELSA-like electron source is selected to maximize electron packet charge with minimal emittance.
The electron bunch radius of 40~$\mu$m is selected to reduce the emittance spectral broadening term in Eq.~(\ref{eq:bandwidth}).
To match the electron bunch radius and pulse duration ($90~$ps), a laser energy of 133~J and focal length of $f_\# = 61$ are required.
The energy and focus are comparable to the BELLA laser (40~J, $f/65$) although that system delivers much shorter pulses (30~fs) \cite{JQE:Nakamura:2017}.
The energy and pulse length are less than an OMEGA EP short-pulse beam (1 to 2~kJ, 100~ps) but focused using a much longer focal length \cite{JPCS:Meyerhofer:2010}.
Note that this design is in the spatial limit of Eq.~(\ref{eq:N_source}) ($z_{\textrm{R}} \ll \tau_{\textrm{L}} c \approx 3$~cm). 
The on-axis design produces $2.9\times 10^{10}$ scattered photons, with a bandwidth of 2.5\%.
If an off-axis laser-electron interaction is required with an impact angle $\phi = 2^\circ$, the number of interacting laser cycles is reduced by $\sim$0.14$\times$ compared to on-axis scattering due to the geometric reduction in the interaction length.

Improving the electron-beam emittance would reduce the bunch radius proportionally with $\varepsilon$.
This would in turn reduce the required laser focal length to match the bunch radius as $\varepsilon$ and the required laser energy as $\varepsilon^2$.
However this would also reduce the interaction length by $f_\#^2$, resulting in less scattering overall.
In general, the number of scattered photons benefits from larger $f_\#$ (for longer interaction distances), which produces larger spots and requires higher laser energy as $E_{\textrm{L}} \propto f_\#^2$.
Overfilling the electron packet with the laser may be beneficial since this maintains the number of scattered photons and the source resolution ($\sigma_{x,\textrm{e}}$), reduces the intensity variation observed by the electron packet, and reduces the difficulty of alignment.
Alternatively, the laser could be focused to a smaller spot than the electron packet, increasing the resolution and relaxing the bandwidth constraint due to $\gamma\sigma_{\theta,\textrm{eff}}$.
However this would reduce the number of electrons available for scattering ($\propto (w_0/\sigma_{x,\textrm{e}})^2$) and the scattering path length.

A calculation for a flying-focus design is also shown in the right column of Table~\ref{table:specs}. 
This design produces comparable scattering performance with a substantially reduced electron bunch charge (100~pC) due to the high intensities ($a_0 = 1$) and long interaction lengths (20~mm).
The spatial resolution is also improved to $\sim$10~$\mu$m and the focal length is reduced to $f_\# = 15$.
However, approximately twice as much laser energy (269~J) is required to create the desired laser focus.
This design requires on-axis focusing due to the co-axial nature of the spatiotemporal pulse shaping.

% For VISRAD model:
% far laser:  2-degrees off axis, f/50
% electron beam:  gamma*sigma_theta = 0.14 rad --> sigma_theta = 0.14/(46) = 3 mrad
% 	f_effective = 1/(2 tan sigma_theta) = 167

\begin{table}[tb]
	\begin{tabular}{|l|c|c|c|c|c|}
		\hline 
		Quantity & symbol & \multicolumn{2}{|c|}{Gaussian laser} & \multicolumn{2}{|c|}{Flying focus} \\
		\hline \hline
		X-ray energy & $\hbar \omega_{\textrm{f}}$  & \multicolumn{2}{|c|}{20~keV} & 10~keV & 50~keV \\
		Electron beam energy & $E_{\textrm{e}} (\gamma)$ & \multicolumn{2}{|c|}{32.8~MeV (65.2)}  & 23~MeV (46) & 52~MeV (103) \\
		\hline
		Collimation angle & $\theta$ &  \multicolumn{2}{|c|}{ 4.2~mrad }  &  6.0~mrad & 2.7~mrad \\
		\hline
		Bunch charge & $Q$  & \multicolumn{2}{|c|}{ 14~nC } &  \multicolumn{2}{|c|}{ 0.1~nC } \\
		Bunch width & $\tau_{\textrm{e}}$  & \multicolumn{2}{|c|}{ 90~ps } & \multicolumn{2}{|c|}{ 1.0~ps } \\
		Emittance & $\varepsilon$  & \multicolumn{2}{|c|}{ 4~mm~mrad } & \multicolumn{2}{|c|}{ 1~mm~mrad } \\
		Bunch radius & $\sigma_{x,\textrm{e}}$ &  \multicolumn{2}{|c|}{ 41~$\mu$m } & \multicolumn{2}{|c|}{ 10~$\mu$m }  \\ 
		Electron bandwidth & $\Delta\gamma/\gamma$ & \multicolumn{2}{|c|}{0.001} & \multicolumn{2}{|c|}{0.001} \\
		\hline
		Laser wavelength & $\lambda_0$  &  \multicolumn{2}{|c|}{1053~nm}  &  \multicolumn{2}{|c|}{1053~nm} \\
		Laser bandwidth & $\Delta\omega_{\textrm{i}}/\omega_{\textrm{i}}$  &  \multicolumn{2}{|c|}{0.001}  &  \multicolumn{2}{|c|}{0.01} \\
		Laser focus & $f/\#$  &  \multicolumn{2}{|c|}{61} & \multicolumn{2}{|c|}{15} \\
		%   &   &    \multicolumn{2}{|c|}{}  & \multicolumn{2}{|c|}{(D = 50 cm)} \\
		Focal spot radius & $w_0$  &  \multicolumn{2}{|c|}{41~$\mu$m} & \multicolumn{2}{|c|}{ 10~$\mu$m } \\
		Rayleigh length & $z_{\textrm{R}}$  &  \multicolumn{2}{|c|}{5.0~mm} &  \multicolumn{2}{|c|}{310~$\mu$m} \\
		Laser duration & $\tau_{\textrm{L}}$ & \multicolumn{2}{|c|}{90~ps} & \multicolumn{2}{|c|}{67~ps (chirped)}  \\
		Peak Intensity & $a_0 (I)$ & \multicolumn{2}{|c|}{0.15 (2.8$\times10^{16}$ W/cm$^2$)}  & \multicolumn{2}{|c|}{1.0 (1.2$\times10^{18}$ W/cm$^2$)}\\ 	
		Laser energy & $E_{\textrm{L}} $  & \multicolumn{2}{|c|}{ 133~J} & \multicolumn{2}{|c|}{ 269~J} \\ 
		\hline	
		Impact angle & $\phi$  &  0~mrad ($0^\circ$) &  35~mrad (2$^\circ$)  &  \multicolumn{2}{|c|}{ 0~mrad (0$^\circ$)} \\
		Interaction length & $L$ & 9.9 mm & {1.4~mm} &  \multicolumn{2}{|c|}{ 20~mm } \\
		\hline \hline
		\# photons & $N_{x,\textrm{tot}}$ & $2.9\times 10^{10}$ & $4.0\times10^{9}$  & \multicolumn{2}{|c|}{ $1.9\times10^{10}$ } \\
		\hline
		Bandwidth & $\Delta \omega_{\textrm{f}} / \omega_{\textrm{f}}$ & \multicolumn{2}{|c|}{ 2.5\%} & \multicolumn{2}{|c|}{ 2.5\%} \\
		\hline
	\end{tabular}
	\caption{Expected performance of a source for x-ray Compton scattering on OMEGA.\label{table:specs}}
\end{table}

%\begin{equation}
%	d = 2z\left[1 + \frac{w_0^2}{z^2}\left(1 + \frac{z^2}{z_{\textrm{R}}^2}\right)\left[\ln{b} - \frac{1}{2}\ln\left(1+\frac{z^2}{z_{\textrm{R}}^2}\right)\right]
%	\label{eq:highI_path}\right]
%\end{equation}

%roughly an ellipsoidal volume, with minor radius $w_0\sqrt{\ln 2} \approx 0.83 w_0$ and major radius $\sqrt{3}z_{\textrm{R}}$.
%$r_{maj} = w_0\sqrt{\ln a};\quad z_{maj} = z_{\textrm{R}}\sqrt{a^2-1}$

\section{Simulation results}
\label{sec:sims}
%%% contributed by: Vlad Musat
Extensive research has been done to develop codes capable of rigorously simulating inverse Compton scattering (ICS) \cite{Krafft2010ComptonRadiation}. 
ICS can be classified in terms of the laser field strength parameter $a_0$ and the recoil parameter $X=4\gamma \hbar \omega_{\text{i}}/m_{\textrm{e}} c^2 \approx 2 \chi/a_0$ \cite{Ranjan2018SimulationLinewidth}. 
If $a_0 \ll 1$, the scattering is linear, i.e., a purely harmonic motion is induced by the external electromagnetic field for the electrons. 
Otherwise, the interaction is nonlinear, which generates higher harmonic modes in the scattered photons. 
For the cases presented here, $X < 10^{-3} \ll1$, such that the electron recoil is negligible and the Thomson regime applies, with a constant cross section $\sigma_T = 8\pi r_{\textrm{e}}^2/3$, where $r_{\textrm{e}}$ is the classical electron radius. 
%Generally, ICS interaction is described by the Compton cross-section. 

%A widely used method to simulate ICS is by using the cross section of the process involved. The Klein-Nishina cross-section \cite{Klein1929UberDirac} is typically used, since it applies to both the quantum linear and nonlinear regimes. 
To assess the validity of the analytical estimates developed in Sec.~\ref{sec:designcalcs}, we have performed simulations of the conditions given in Table~\ref{table:specs} using the code \textit{RF-Track} \cite{RFT_manual}.
\textit{RF-Track} is a fast and parallel Monte Carlo-based particle tracking code developed at CERN that includes the option to compute ICS interaction using the Klein--Nishina cross section. 
\textit{RF-Track} has been recently benchmarked \cite{Musat2022ATechnology} against \textit{CAIN} \cite{Chen1995CAIN:Non-lineaires}, the standard Monte Carlo code used to simulate ICS in the linear scattering regime.
\textit{CAIN} includes physics covering both the linear and weakly nonlinear regime in the classical and quantum domain, including the physics of collision angle, multiple scattering, and the polarisation of scattered photons, and has been extensively benchmarked against experimental results from ICS sources \cite{Sun2011TheoreticalSource}. 
%A code snippet of \textit{RF-Track} in Octave is presented in Listing~\ref{listing:RFTsnippet}.
The Gaussian laser configuration discussed above has an amplitude $a_0 = 0.15$, which places the interaction in the linear regime, suitable for \textit{RF-Track}. 

\begin{figure}
\includegraphics[width=0.6\textwidth]{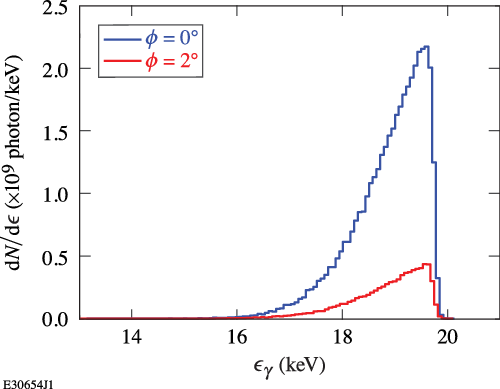} %{photon_spectra_gaussian.png}
\caption{X-ray photon spectra of the Gaussian laser configuration, with crossing angles of 0$^\circ$ and 2$^\circ$. The number density of scattered photons is plotted against the scattered photon energy.\label{fig:RFT_spectra}}
\end{figure}

Scattered photon spectra computed in \textit{RF-Track} for the Gaussian laser configurations given in Table~\ref{table:specs} are shown in Figure~\ref{fig:RFT_spectra}. 
The Compton edge and scattered photon bandwidth are similar for both crossing angles, indicating a weak dependence of the scattered photon energy on this parameter. 
The number of scattered photons significantly increases for the on-axis collision, as expected given the increased scattering length.

Parameters of the scattered photon spectrum simulated in \textit{RF-Track} are included in Table~\ref{tab:RFT_results}. 
The scattered photon energy is slightly smaller than the expected value. 
This can be accounted for by the additional effects in \textit{RF-Track}, which lead to a decrease in the Compton edge, such as a nonzero recoil and $a_0$, which were assumed for the theoretical estimate. 
The analytical predictions of the number of photons generated per interaction and the bandwidth of the scattered photons are closely matched by the simulation.  %was slightly overestimated by the simulation. 
This result supports the analytical model presented in Sec.~\ref{sec:designcalcs}.
%A small difference is typically expected when comparing simulation results with theoretical ones. 
%The bandwidth of the scattered photons matched the theoretical prediction.

%\begin{listing}[]
%\begin{minted}{octave}
%RF_Track;
%%% Create the bunch
%B0 = Bunch6d(RF_Track.electronmass, Q, -1, Pref, Twiss, N_macroparticles);
%%% Define laser-beam IP region
%FP = LaserBeam(); 
%%% Define lattice
%L = Lattice();
%L.append(FP);
%%% Tracking
%B1 = L.track(B0);
%%% Post-processing
%M1 = B1.get_phase_space('%x %xp %y %yp %t %Pc %m %N');
%\end{minted}
%
%\caption{Code snippet of \textit{RF-Track} in Octave, used to simulate the ICS interaction.}
%\label{listing:RFTsnippet}
%\end{listing}

\begin{table}
\caption{Parameters of scattered photons generated using \textit{RF-Track}. The mean values and errors were determined from ten runs of the simulation code.\label{tab:RFT_results}}
\begin{tabular}{|l|c|c|c|}
\hline
Quantity & Symbol  & \multicolumn{2}{c|}{Gaussian laser}\tabularnewline
\hline
\hline
Impact angle & $\phi$ & 0 mrad (0$^\circ$) & 35 mrad (2$^\circ$)\tabularnewline
Compton edge & $\hbar\omega_{\textrm{f}}$ & 19.47$\pm$0.06 keV & 19.46$\pm$0.08 keV\tabularnewline
\# photons & $N_{x,\textrm{tot}}$ & (3.13$\pm$0.01)$\times10^{10}$ & (6.23$\pm$0.01)$\times10^{9}$\tabularnewline
Bandwidth  & $\Delta\omega_{\textrm{f}}/\omega_{\textrm{f}}$ & (2.62$\pm$0.12)\% & (2.66$\pm$0.11)\%\tabularnewline
\hline
\end{tabular}
\end{table}

\section{Implementation at OMEGA}
\label{sec:implementation}
Integrating this novel x-ray source with an existing high-power HED facility introduces several novel constraints on the design of the system.
In this section we consider several of the design challenges that must be overcome to implement such a source at the Omega Laser Facility.

The simplest approach to integrating an ECOS x-ray source with the OMEGA-60 or OMEGA-EP target area would locate the electron acceleration stages and laser interaction chamber next to the OMEGA target chamber with a fixed port location.
In this design, the electron beam is dumped outside of the OMEGA target chamber, and only a collimated x-ray beam is injected into the chamber.
A benefit of this design is the ability to dump the electron beam and scattering laser far from the sensitive diagnostic instruments around the target chamber.
However, this design limits the x-ray flux on the laser-driven target, as the scattering event would occur several meters from TCC and the x rays diverge from the interaction point.
To achieve a collimation of $\theta\gamma \lesssim 0.27$~radians as described in Table~\ref{table:specs}, photons of 10 (50)~keV [$\gamma = $~46 (103) for a 1~$\mu$m wavelength laser] require collimation angles of 5.9 (2.6)~mrad, respectively.
If the interaction occurs 3~m from target chamber center (twice the OMEGA target chamber radius), the collimated beam would then project to a radius of 18 (8)~mm at TCC: almost an order of magnitude larger than a typical target.  
Increased collimation reduces the bandwidth at the cost of signal: a 1~mm beam at TCC would require collimation of 0.33~mrad and collect less than 0.1\% of the scattered signal.

This problem could be mitigated by the use of x-ray optics to collect and collimate x-rays of a desired wavelength to the experimental chamber.
Issues when considering the use of x-ray optics in this application are cost, complexity, and efficiency.
High-efficiency ($>$80\%) lenses have been demonstrated for $>$10~keV x-rays using a multi-layer Laue geometry \cite{LSA:bajt:2018}.
Such lenses typically suffer from chromatic aberration, limiting their use to a specifically designed narrow-band wavelength and reducing the absolute efficiency for sources with bandwidth.  
Achromatic lenses have also been demonstrated using a combination of optics, but with reduced efficiency \cite{NC:Kubec:2022}.
Because the proposed single-shot x-ray source is primarily limited in the number of x rays produced, the efficiency of the source is paramount to this approach.
We therefore consider methods to reduce the distance from scattering point to the target area 
\footnote{Increased collimation may also be achieved by using a long wavelength scattering laser, compensated by higher electron beam energy.  
A CO$_2$ laser ($\lambda = 10~\mu$m) would produce 10-50~keV photons by scattering from an electron beam with $\gamma = $142-318.  
Optimal collimation would then capture 1.9-0.85~mrad, projecting to 5.7-2.5~mm at TCC, respectively.
This is somewhat improved over the prior case, but the illuminated area is still larger than most targets.
Moreover, this comes at the cost of a considerably larger and higher-energy electron gun.
}.

Assuming a 1~$\mu${m} laser is used, to achieve a beam radius less than 1~mm for photons above 10~keV would require the scattering to occur roughly 15~cm from TCC.
This concept would require that the electron beamline is injected into the OMEGA target chamber at a fixed port location.
Final beam steering and shaping magnets would point the beam to TCC and control its focus.
In this arrangement, the scattering laser cannot be injected directly opposite the electron beam, as the target is in the way.  
Three options are available.
With the final optics of the scattering laser on the opposite side of the target chamber,  either a non-zero incidence angle $\phi$ would be introduced to avoid TCC, or the scattering laser may be apodized to prevent striking the target.
Third, the final optics may be positioned co-linear with the electron beam on the near side of TCC.  
These cases are considered below.

\begin{figure}
	\includegraphics[width=\textwidth]{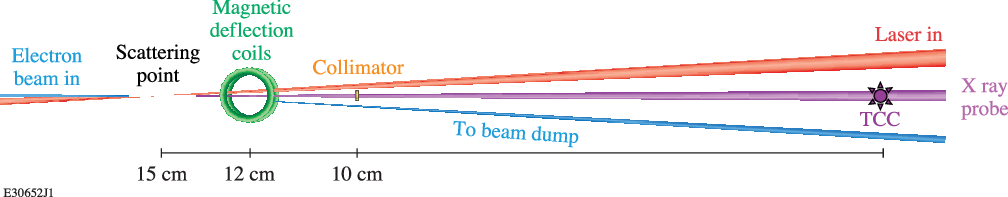} %{Visrad_farConcept.png}
	\caption{Cartoon of laser-electron interaction region for case with the final laser optic on the opposite side of TCC from the electron beam.  This image assumes $\phi = 2^\circ$ and electron deflection angle 3$^\circ$ ($B\sim1$~T).\label{fig:faroptic}}
\end{figure}

If the laser final optic is opposite TCC from the electron beam entrance port, an incidence angle of $\phi \approx 0.033$~rad $(2^\circ)$ and a beam focus $f_\# > 1/\tan(2\phi) \sim 15$ would provide a 5~mm standoff from the target hardware at TCC. %\sim \tan^{-1}\left(5\textrm{mm}/150\textrm{mm}\right) 
Following Eq.~\ref{eq:N_source}, the scattered photon number would increase quadratically with focal length up to $f_\# = 20$, and linearly above that. 
This arrangement has the disadvantage that plasmas near TCC may perturb the beam transport, and that the quadratic increase in scattering volume with longer focal lengths cannot be leveraged.
This scenario is depicted in Figure~\ref{fig:faroptic}.

If the laser final optic is opposite TCC from the electron beam and apodized to avoid target hardware near TCC, similar calculations require the apodization to subtend at least $2^\circ$.  
However, this places an upper limit on the f-number of the final optic: the apodized beam is limited to a focal length of f$_\# <$ 15, whereas a short focal length is undesirable for this application.
This requirement will be further reduced by the need to avoid a collimator foil.  
As such, apodization of an on-axis opposing beam is not likely to succeed for the Gaussian-beam application.
However, for the case of a flying-focus laser, the interaction length is decoupled from the focal length and this approach may succeed.
Within the OMEGA target chamber, an f/2 OAP is currently used to focus the OMEGA-EP short-pulse beam during joint operations
Using a comparable optic, apodization of 11\% of the beam area would allow a 5~mm offset for the scattering laser from all sides of a target and stalk positioned at TCC.

\begin{figure}[hb]
	\includegraphics[width=\textwidth, trim={0, 80, 0, 0}]{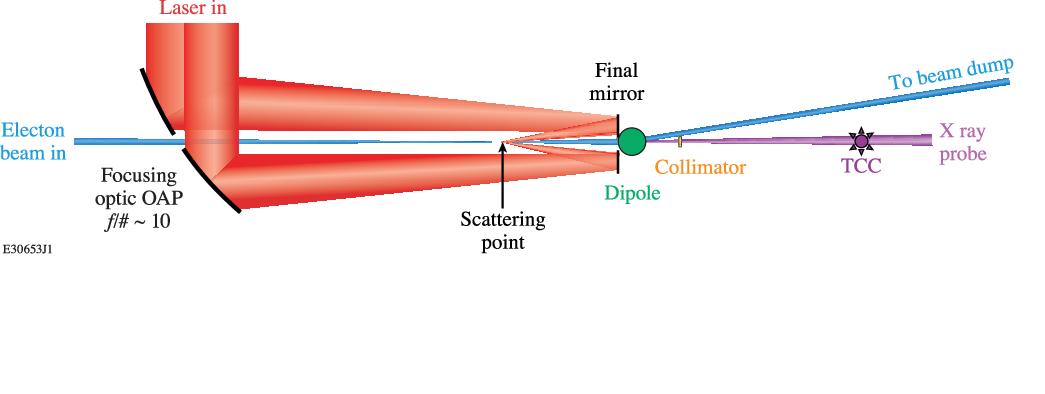} %{Cartoon_nearConcept.png}
	\caption{Cartoon (not to scale) of laser-electron interaction region for case with the final laser optic on the same side of TCC as the electron beam.\label{fig:nearoptic}}
\end{figure}

If the final optic is positioned on the electron beam axis prior to TCC, its location must take into account the electron beam dump magnet and the x-ray collimator.  
To use the system on cryogenic target implosions, all hardware must be at least 10~cm from TCC on an equitorial line of sight (the distance increases for non-equitorial views).  
The maximum distance between the final optic and the scattering location is then $D = 5$~cm.
Intensity on this final optic will scale as the intensity at best focus times a geometric ratio $R = f_\#^4 \lambda^2/D^2$.
For $D = 5$~cm, $\lambda = 1~\mu$m, and $f_\# = 10$, this ratio is $4\times10^{-6}$ and an $a_0 = 0.15~(I = 2.8\times10^{16}~\textrm{W/cm}^2)$ beam will produce an intensity of $10^{11}$~W/cm$^2$ on the mirror.
This intensity is approaching the threshold for optic damage, so use of a plasma mirror for this final stage may be required.
A plasma mirror is capable of reflecting light at above TW/cm$^2$ intensity, which would enable placing the final mirror closer to the scattering point and using longer focal lengths.
%A shaped convex plasma mirror could potentially be used to upconvert the $f_\#$ of the final beam, achieving much longer interaction volumes than are otherwise possible.
A cartoon of this scenario is depicted in Figure~\ref{fig:nearoptic}.
Because flying-focus laser intensity is generally elongated along the axis of the final optic, a co-axial geometry would be needed for a flying-focus based source.

% Total Energy = integral(intensity)*dt*dz*(2pi r)dr
% For Gaussian beam, Energy = 

A magnetic deflection system between the scattering region and TCC would steer the electrons away from TCC to a beam dump on the opposite side of the chamber.
A collimator would also be needed to block non-monochromatic photons produced at larger scattering angles.
Challenges of this scenario include co-timing of the electron beam and scattering laser, and alignment of the beam, laser, and collimator.

%Given typical gradients and the relatively modest energies of interest (up to 50~MeV), a custom accelerator might be designed to fit within the space limitations of a TIM.
%However, engineering challenges, such as transporting short-pulse laser light to the photocathode and rf waves into the acceleration cavity, might prove prohibitive.

\subsection{Beam Dump Requirements}
To prevent the electron beam impacting the experiment at TCC, the electrons must be deflected to a beam dump.   
A magnetic dipole field produced by a capacitor discharge through a magnetic field coil may be fielded between the scattering region and TCC, similar to the magneto-inertial fusion electrical discharge system (MIFEDS) that has been implemented for magnetized plasma experiments on OMEGA \cite{PRAB:Shapovalov:2019}.
The deflection must occur prior to x-ray collimation because if the electrons were to strike the high-$Z$ collimator foil, this would produce a large, broadband bremsstrahlung source that would likely overwhelm the Compton scattering signal.
%The gyroradius of 50~MeV electrons is approximately 17~cm in a 1~Tesla magnetic field.
The angle of deflection $\zeta$ for relativistic electrons traversing a magnetic field is given by the scaling formula $\sin\zeta \approx (BL/3.33~\textrm{T}~\textrm{cm})(E_{\textrm{e}}/10~\textrm{MeV})^{-1}$.
A deflection of 3$^\circ$ would avoid striking the experiment from a distance of 10~cm, requiring a magnetic field integral of at least 0.9~T~cm.
This is readily achievable using MIFEDS-3, which has demonstrated peak fields over 30~T in an 0.8~cm region \cite{PRAB:Shapovalov:2019}.
For electron beams at lower energy, the coil field can be detuned to ensure the electron beam reaches a beam dump located on nearly the opposite side of the target chamber wall from the electron source.

\subsection{Collimation Requirements}
Collimation of the x-ray source is needed to achieve narrow bandwidth, as shown in Fig.~\ref{fig:analyticmodel}(b).
The $e$-folding attenuation depth for 50-keV x rays in tungsten (tantalum) is 87 (105) $\mu$m, respectively.  
Attenuation to $<$1\% of the signal can therefore be achieved by a 0.5-mm-thick foil fielded between the source and TCC at a distance $d$ from the source.
The radius of the collimating aperture is required to be at most $r_a = d\tan\theta_{\textrm{max}} \approx 0.27 d/\gamma$ for 2\% bandwidth, and becomes smaller as the electron beam energy increases.
For the highest energy x rays, $\gamma \approx 100$, the required aperture radius scales as $r_a = 27$~$\mu\textrm{m}\times\left(d/\textrm{cm}\right)$.

Co-alignment of the electron source, aperture, and TCC must be achieved to on the order of the collimator radius.  
This requirement becomes easier as the collimator distance increases.
Assuming the electron beam repetition rate is of the order of 1~Hz, alignment of the electron beam and collimator may be achieved either by operating the full scattering source at high-repetition rate and low power, or by directly irradiating a scintillator or phosphorescent screen at TCC using electrons transiting the collimator.  
In general, a high repetition rate mode for the electron beam and laser will be beneficial for fine-tuning the alignment, co-timing, collimation, beam energy, and beam deflection, prior to operating at full power on the integrated HED experiments.

%%% Ideas from Gerrit:  Feb 4 2021
% 1. Colliders use a solenoid to pinch the electron beam and get down to a small focused spot.  He gestured a ~1m long solenoid with peak focus in the middle somewhere.  Could do this with a MIFEDs solenoid to pinch the beam and get a small electron focal spot.
% 	UPDATE: this is discussed in the Kramer 2016 paper on the electron beam final focus system for Thomson scattering at ELBE.  They predict (achieve?) for an emittance of 13 mm mrad and gamma = 50 a focal spot size of 40 um RMS with divergence of 10 mrad using the FFS.  Note that 50*0.04*10 = 20 mm mrad, so there's some imperfection in the focusing, but not a lot.  This is probably mostly due to beam energy spread (0.01 in the prediction). 
%
% 2. Could use a storage ring to take advantage of some improved beam quality tricks.  Mentioned bunching using a cavity: take several individual bunches and compress them in time to a single macro-bunch that is injected to the interaction point.  Might be able to get ~ 100's nC this way.  People expermienting with this:  the IOTA storage ring at Fermilab's FAST accelerator.  
%
%%%%%

\section{Applications}
\label{sec:apps}
The source described above will have high utility as a probe for a variety of HED experimental platforms and conditions.
Here we consider its applications in x-ray diffraction, inelastic x-ray scattering, x-ray absorption fine-structure measurements, and imaging applications.

\subsection{X-Ray Diffraction}
\begin{figure}
    \includegraphics{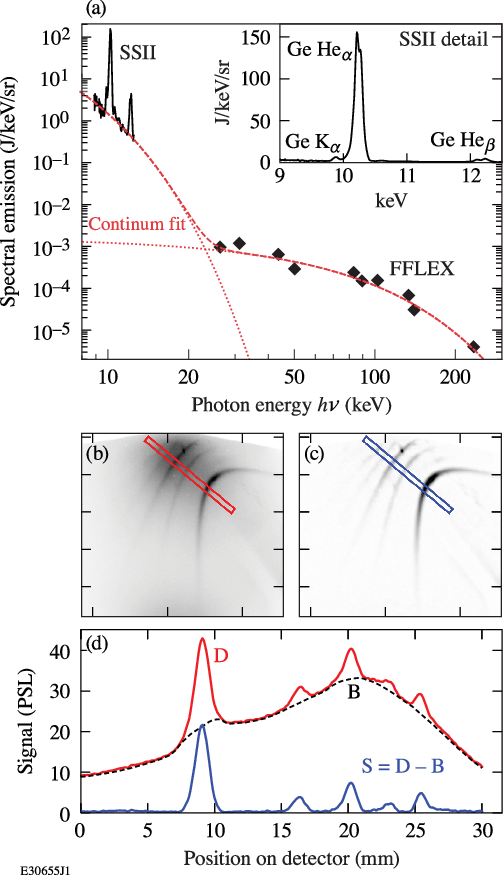} %{Rygg_combined.pdf}
    \caption{X-ray diffraction data collected on the NIF.  (a) The 11-keV x-ray line of interest (Ge He$_\alpha$) sits on a two-exponential background containing a similar amount of energy ($E_{\textrm{BG}} / E_{\textrm{line}} \approx 80\%$). (b) The raw data require (c) background subtraction to obtain (d) analyzable diffraction lines.  The signal-to-background ratio is often below unity, obscuring small peaks and detail in the signal.  Images reproduced with permission from Ref.~\cite{RSI:Rygg:2020}.\label{fig:diffraction}}
\end{figure}

The present state-of-the-art x-ray diffraction experiments on OMEGA and NIF typically use laser-driven metal foils as He$_\alpha$ x-ray backlighters \cite{RSI:Rygg:2020,RSI:Rygg:2012,Polsin2022,Nature:Millot:2019,RSI:Coppari:2019}.
These sources are capable of producing monoenergetic x rays up to roughly 10~keV.
In order to maximize x-ray production, up to 2-ns laser pulses are used to drive the backlighters.
This limits the structural determination to only simple crystal systems, inhibits the ability to explore phase transformation kinetics, limits drive pressure because of competing background x rays, and limits applicability to materials compressed by comparably long laser pulses.  %and prevents their use on spherically imploded targets that can reach higher pressure states.%
Under these conditions, Fe backlighters can probe the samples with 10$^{12}$ photons per experiment \cite{RSI:Rygg:2020}; however, efficiency decreases with increasing atomic number such that higher energies cannot be effectively used.
Radiation from the backlighter and x-ray sources also produce substantial background on the detector at and above the energies of interest that must be subtracted to extract the signal.
The high laser intensities needed to produce efficient He$_\alpha$ line radiation also produce hot electrons by laser plasma interaction physics that result in a broad, hard x-ray background.
An example of Ge He$_\alpha$ x-ray diffraction data collected on the NIF using 26 kJ to drive the backlighter is shown in Fig.~\ref{fig:diffraction} \cite{RSI:Rygg:2020}.
The signal-to-background ratio in the collected data is below unity for many of the recorded peaks.  
The need to subtract the background in order to analyze these peaks introduces uncertainty and obscures details in the diffraction measurement. 
 
An ECOS backlighter would improve x-ray diffraction studies by introducing higher-energy x rays (20 to 50~keV) and improved time resolution ($<$100~ps).
Use of the Compton scattering source would eliminate the hot-electron background associated with the x-ray source foil.
Sampling with x rays above 20~keV would benefit the measurement in three ways: by allowing an increase in the detector shielding to reduce x-ray background from the driven sample $(T \sim $ few keV); by reducing the x-ray scattering dispersion for a given lattice spacing, which results in higher x-ray fluence in the signal region; and by increasing the number of accessible scattering planes (Q-range) in the sample.
These improvements are expected to  reduce the background by more than two orders of magnitude and allow for the determination of complex crystal and liquid structures.
This benefit would compensate for the reduction in scattering signal; however, at least $10^{10}$ photons in the source would likely be required.

\subsection{Inelastic X-Ray Scattering}
% G. Gregori
% Probably a good reference:  
Inelastic X-ray Scattering (IXS) has been an important diagnostic for experiments at laser facilities for many years \cite{Glenzer2009}.
In this technique, x-ray scattering spectrally resolves material excitations from both electron plasma waves and ion-acoustic oscillations, and the resulting spectral shape and dispersion provide information that can be used to infer the equation of state of the plasma as well as transport properties.
Presently, electron plasma waves are primarily used to investigate dense matter states. 
These modes are separated by a few tens of electron volts and can thus be resolved with well-established methods \cite{Glenzer2009}, such as crystal spectrometers. 
While x-ray probe beams from FELs operating in seeded mode can achieve spectral bandwidth of $\sim$1~eV at 10~keV (0.01\%), experiments on laser facilities have been limited to x-ray sources produced in the same manner as described above for diffraction experiments. 
These line-radiation sources impose severe limitations in terms of spatial and temporal coherence \cite{Gregori2006} and effectively limit what information can be extracted from the data. 
IXS experiments using line-radiation sources can at most resolve the plasmon peaks \cite{Glenzer2007}, but only in strongly driven samples.
Extracting dynamical properties (i.e., the collision frequency) from the width of those peaks is challenging. %(see Figure~\ref{ixs1} for an example from Ref.~\cite{Glenzer2007}). 
On the other hand, if data can be collected at sufficiently high spectral resolution, as in experiments at FEL facilities, then the plasmon peaks or even ion-acoustic peaks can be well resolved \cite{Fletcher2015,McBride2018}, and further information on transport and dynamics become accessible.

%% Notes HGR 1/30/23
% The relevant comparison points are: # photons from the source interacting with the sample, and bandwidth of the source.
% From Glenzer2007:
% # Photons: "we estimate $10^{14} Cl Ly_\alpha$ x-ray probe photons at the sample. Together with the cross section for collective scattering and scattering length of ell = 400 um we obtain a scattering fraction of \sigma n_e ell = 2x10^{-3} and 2x10^{11} scattered photons for single-shot scattering experiments."  --> 1e14 total at sample. Presumably similar detectors would be used with ECOS.
% Bandwidth: from Fig. 7a (inset) appears to show FWHM ~ 10 eV --> 0.3% bandwidth.  
% comparing Table 1: ECOS cannot compete on either of these fronts.  Why the difference from diffraction?  Perhaps because the IXS work has used lower energy (~2-3 keV) so there's more photons at the same efficiency.  ALso because the gain with ECOS is reduced background.  Does reduced background help us here?

%\begin{figure}
%    \includegraphics[width=1.0\textwidth]{FigIXS1.png}
%    \caption{IXS experiments on OMEGA. (a) Experimental setup for collective x-ray forward scattering is shown along with the scattering vector diagram and the x-ray probe spectrum. (b) Experimental scattering data are shown along with the theoretical fit. (c) Comparison of the plasmon spectrum with calculations for three different densities. Reproduced with permission from Ref.~\cite{Glenzer2007}.\label{ixs1}}
%\end{figure}

The ECOS source that has been proposed here cannot directly compete with line-radiation sources as those can still produce a larger x-ray fluence, nor with FELs that can provide narrower bandwidths. 
However, where ECOS becomes competitive is in accessing higher x-ray energies. 
At energies above $10$~keV, line emission becomes much less efficient. 
This is strikingly evident in IXS applied to probing the conditions in the in-flight DT-ice layer in ICF implosions on OMEGA, as described in Ref.~\cite{POP:Poole:2022}.
The x-ray energies in that study are lower than those considered here (2--3.5 keV), and the bandwidth was assumed to be less than 0.5\% (10~eV). A marginally diagnosable signal was obtained with an x-ray fluence of 2.5$\times$10$^{13}$~photons/sr, or 2.7$\times$10$^{11}$ photons interacting with the target.
This produced an estimated 3$\times$10$^{7}$ scattered photons, with 300 ultimately detected.   

\begin{figure}
\includegraphics[width=\textwidth]{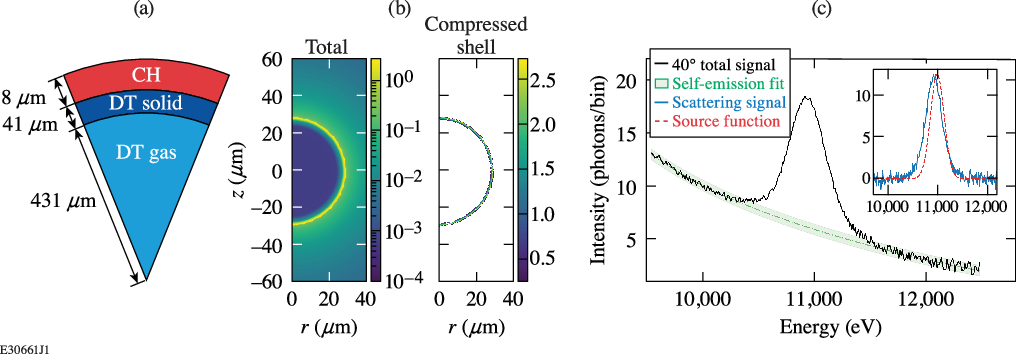} % {Figure2_ECOS.png}
    \caption{(a) Simulated target design, with an in-flight adiabat of 5.4. (b) 2-D mass density conditions in the ICF implosion at two-thirds compression ($t = 1530$ ps) from DRACO simulations. The region of the compressed DT shell is highlighted. (c) Total detected signal per bin, assuming $10^{-5}$ detector efficiency and a bin size of $10$ eV, integrated over the x-ray pulse.  See Ref. \cite{POP:Poole:2022} for details on synthetic IXS analysis. \label{fig:ixs2}}
\end{figure}

The ICF case is significantly more challenging than other WDM plasmas of interest, due to the low electron density in hydrogenic fuels and the large background emission. 
A higher photon energy source becomes valuable to penetrate the denser material and increase the signal-to-background ratio.
Figure~\ref{fig:ixs2} presents a reassessment of the ICF IXS case assuming an ECOS-generated 0.1~mJ, 90~ps x-ray pulse generating 11~keV X-rays with 275~eV bandwidth.
A 50~$\mu$m spot size incident on the in-flight capsule and a scattering angle of 40$^{\circ}$ was assumed. 
(See Ref.~\cite{POP:Poole:2022} for details on the synthetic IXS analysis.)
The higher x-ray energy results in both reduced absorption in the target and a lower value of the scattering parameter $\alpha = 1/k\lambda_{De} \approx 0.2$.
This results in a larger probability of scattering and a reduced background, although the scattering is consequentially far into the non-collective regime.
Approximately 500 scattered photons were detected, of which 120 were scattered inelastically.
The predicted spectrum is encouraging as it shows measurable differences compared to the incident beam profile. 
As such, an ECOS x-ray source provides some utility for IXS as a diagnostic of compressed ICF capsules.

\subsection{X-Ray Imaging}

\begin{figure}
    \includegraphics{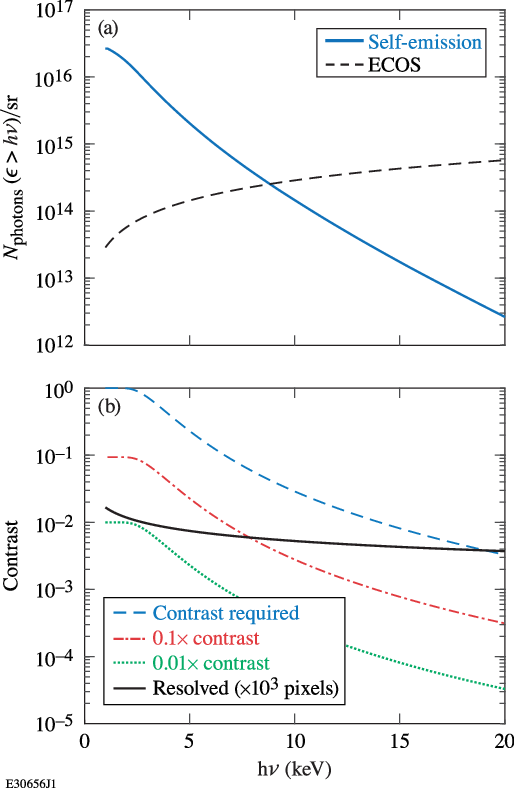} %{Compton_ICF_comparison.png}
    \caption{(a) Comparison of the x-ray fluence produced by an ICF implosion on OMEGA calculated from Ref.~\cite{POP:Cao:2019} (blue) with fluence from the electron-beam Compton source described in Table~\ref{table:specs} (black dashed).  (b) Contrast required to resolve the darkest feature in a backlit image (blue), and 10\% (red), 1\% (green) of that value, compared with the statistical resolution of the described source assuming photon statistics and a 1000-pixel image (black).\label{fig:imaging}}
\end{figure}

The requirements of x-ray imaging are quite different from diffraction and IXS.
For imaging, beam divergence is beneficial since it produces magnification of the image and simplifies diagnosis.
For point-projection imaging, the resolution would be set by the size of the source, which is the smaller of the electron packet width and the scattering laser focal width.
Given the increase in bandwidth with reduced electron packet radius, for imaging applications it is best to allow increased bandwidth in favor of improving the spatial resolution.
Consider a case in which the electron beam is focused to a small point $\sigma_x \sim 1~\mu$m.
In this case, it will have a divergence $\sigma_\theta = \varepsilon/\beta\gamma\sigma_x \approx50$~mrad for an ELSA-like beam, dominating the x-ray divergence ($\theta \sim 1/\gamma \sim 15$~mrad).
This increased divergence has two effects.
First, the field of view at the target plane is increased to roughly 9~mm: this is much larger than the typical target.
Second, the fluence at the target plane is reduced proportionally to $\sigma_\theta^{-2}$.  
In estimating x-ray imaging applications, we must therefore trade off resolution with photon statistics.

We consider here the requirements for x-ray backlighting of an imploded inertial confinement fusion (ICF) target at peak compression.
This measurement has not been successfully performed on OMEGA due to the bright self-emission of the imploded target and the small spatial resolution required.
The self-emission of an OMEGA cryogenic ICF implosion is characterized by a roughly thermal distribution, with temperature in the range 2.8 to 3.5~keV and total emission of roughly 8~J/sr \cite{POP:Cao:2019}\footnote{Total x-ray yield in experiments is typically 1/3 of the simulated values plotted in the reference.}.
The self-emission photon fluence above a given energy is shown in Fig.~\ref{fig:imaging}, in comparison with a Compton source fluence calculated using Eq.~(\ref{eq:N_source}) and the small angle approximation, $\Omega \approx \pi \theta^2$.  
The source parameters are taken from Table~\ref{table:specs}.
We observe that the fluence of the described source exceeds the self-emission fluence from the implosion for a source energy of approximately 10 keV and above, as shown in Fig.~\ref{fig:imaging}(a).

At higher photon energy, the opacity of the compressed target is reduced.
This sets a limit for the resolved contrast that is required to record an image, as shown by the blue curve in Fig.~\ref{fig:imaging}(b).
For example, at 10~keV, the most opaque limb of the reference implosion absorbs only 2.9\% of the x rays, and a backlighting source must resolve this perturbation.
Contrast resolution depends primarily on the number of photons recorded per imaging pixel: with $N$ photons recorded, a statistical uncertainty of $\sigma_N = N^{1/2}$ is expected.
This formula is used to estimate the contrast resolution of the described source as a function of photon energy, as shown by the black line in Fig.~\ref{fig:imaging}(b).
We find that, assuming a 1000-pixel image, the described source is able to resolve the contrast level required to record an image of the imploded target.

Since a narrowband spectrum is not required, an alternative option would be to directly irradiate a high-$Z$ foil with the electron beam.
This will produce a broadband intense bremsstrahlung backlighter, with x-ray energy of hundreds of keV. 
Assuming 1\% energy conversion into x rays, such a source would produce at least an order of magnitude more photons than the Compton scattering source described here, and might provide an alternative if greater contrast is required.  %Is a Geant simulation of this worth adding? I have several built for this very job. The conversion efficiency can also be much higher than 1%, but the photon energy will be in the MeV range G.B.%

\subsection{X-ray Absorption Fine-Structure}
The x-ray absorption fine structure (XAFS) is sensitive to details of interatomic spacing \cite{Koningsberger:1988} and has been used to record changes in the crystal phase and temperature of laser-compressed materials \cite{PRL:Yaakobi:2004, PRL:Yaakobi:2005, PRL:Dorchies:2011, Ping2013}.
Research on the OMEGA laser has used implosions of CH shells to provide an intense and spectrally smooth subnanosecond pulse of x-ray radiation for XAFS measurements \cite{POP:Chin:2022}.
As described above, the ECOS source is predicted to produce higher photon fluence compared to an ICF implosion, especially at energies above 10~keV.  
XAFS measurements require a smooth x-ray spectrum covering a region near the x-ray absorption line of the material: for example, the extended XAFS (EXAFS) signal in iron occurs in the range 7.1 to 7.6 keV, requiring a bandwidth of at least 7\% \cite{PRL:Yaakobi:2005}, roughly triple the values described in Table~\ref{table:specs}.
From Eq.~(\ref{eq:bandwidth}), increased bandwidth can be obtained from the ECOS source by increasing the beam divergence $\sigma_{\theta,\textrm{eff}}$ of the electron beam. 
This can be achieved at fixed emittance by increasing the focus of the electron packet using magnetic optics.
Alternatively, this bandwidth could be obtained by increasing the intensity of the laser $a_0$ by a factor of 3$\times$, which would have the additional benefit of increasing the photon number by approximately 9$\times$. 
The ECOS system would have several benefits over implosion backlighters for this research: tunability of the photon energy to match lines of interest; efficiency due to the collimated nature of the source; and the range of accessible conditions since all the OMEGA beams would be available for preparing the sample, rather than driving the backlighter.
%XANES

\subsection{Electron Radiography}
The electron source described above for Compton scattering may also be used directly as a source of probing radiation for HED targets. 
Charged particle radiography with protons is a mature technique at the Omega Laser Facility using laser driven sources \cite{Science:Rygg:2008} and at Los Alamos National Laboratory (LANL) using an 800~MeV linear accelerator. 
Compared to protons, electrons are more penetrating at a given energy and are more sensitive to electric and magnetic fields \cite{LPB:Merrill:2015}. 
%This allows for a probe that penetrates materials of higher areal density than typical proton sources, while also being more sensitive to fields. 
Additionally, electrons can take advantage of magnetic optics to achieve 1-$\mu$m radiography resolution or better. 
Current electron radiography research at Omega is focused on using laser wakefield acceleration-derived electron beams \cite{SR:Bruhaug:2023}, but the low-emittance, monoenergetic beam of a linear accelerator is much better suited to this task and has already been shown to work for static targets with electrons and static and dynamic targets with protons \cite{LPB:Merrill:2015}.

\section{Conclusions}
We have described the requirements for a single-shot electron-beam--based Compton-scattering (ECOS) x-ray source capable of producing at least $10^{10}$ x rays in less than a nanosecond.
The physics of Compton scattering implies several important design constraints.
Narrow bandwidth requires source collimation (less than about 6 mrad), a small electron beam emittance ($\varepsilon \sim$ a few mm-mrad), and a laser amplitude held below $a_0 \lesssim 0.15$.
The divergence of the source requires that the scattering occur in close proximity to the probed experiment (about 17 cm).
Taking into account these considerations, and on the basis of electron sources described in the literature, designs that produce $10^{10}$ scattered photons in the energy range of 10 to 50 keV and with a bandwidth of less than 3\% are technically feasible.
Simulations using the \textit{RF-Track} code closely confirm the analytical results assuming scattering of a matched Gaussian laser pulse.
Additionally, the use of a spatiotemporally controlled (flying-focus) laser with the proposed electron beam has the potential to dramatically increase the number of scattered photons per electron beam charge.

If implemented on the OMEGA or OMEGA EP lasers, this source would greatly extend the sensitivity of present efforts in x-ray diffraction and x-ray near-edge absorption measurements.
Its brightness is predicted to be sufficient for recording radiographs of cryogenic-DT-filled ICF implosions on the OMEGA laser.
In summary, the development of this source would lead to significant and novel results in HED physics over the next decade.

This material is based upon work supported by the Department of Energy National Nuclear Security Administration under Award Number DE-NA0003856, the University of Rochester, and the New York State Energy Research and Development Authority. 
The support of DOE does not constitute an endorsement by DOE of the views expressed in this paper.

This report was prepared as an account of work sponsored by an agency of the U.S. Government. 
Neither the U.S. Government nor any agency thereof, nor any of their employees, makes any warranty, express or implied, or assumes any legal liability or responsibility for the accuracy, completeness, or usefulness of any information, apparatus, product, or process disclosed, or represents that its use would not infringe privately owned rights. 
Reference herein to any specific commercial product, process, or service by trade name, trademark, manufacturer, or otherwise does not necessarily constitute or imply its endorsement, recommendation, or favoring by the U.S. Government or any agency thereof. 
The views and opinions of authors expressed herein do not necessarily state or reflect those of the U.S. Government or any agency thereof.

\appendix
\section{Derivations}
\label{app:derivation}
%In the particle's rest frame, satisfying conservation of momentum requires:
%\begin{equation}
%	\lambda_f = \lambda_i + \frac{h}{m_{\textrm{e}} c}\left( 1 - \cos\theta\right),
%\end{equation}
%\noindent for final and initial wavelengths $\lambda_{f,i}$ and scattering angle $\theta$.

\subsection{Photon Energy and Scattering Probability in the Laboratory Frame}
\label{app:derivation:xs}
The differential cross section for scattering in the rest frame of the electron is given by the Klein--Nishina formula \cite{Klein1929UberDirac}:
\begin{equation}
	\frac{\textrm{d}\sigma_{\textrm{KN}}}{\textrm{d}\Omega} = \frac{r_{\textrm{e}}^2}{2}\left(\frac{\omega_{\textrm{f}}}{\omega_{\textrm{i}}}\right)^2\left(\frac{\omega_{\textrm{f}}}{\omega_{\textrm{i}}} + \frac{\omega_{\textrm{i}}}{\omega_{\textrm{f}}} - \sin^2\theta_{\textrm{s}}\right),
	\label{eq:KleinNishina}
\end{equation}
\noindent where $r_{\textrm{e}}$ is the classical electron radius and $\theta_{\textrm{s}}$ is the scattering angle of the photon.
%MATLAB code: 
%xs_KN = @(boost, angle) 0.0794078766/2*boost.^2.*(boost + 1./boost - sin(angle).^2);  % barns
Notably, the ratio of final to initial frequency is determined entirely by scattering angle and incident photon energy
\begin{equation}
	\frac{\omega_{\textrm{f}}}{\omega_{\textrm{i}}} = \frac{1}{1+\frac{\hbar\omega_{\textrm{i}}}{m_{\textrm{e}}c^2}\left(1-\cos\theta_{\textrm{s}}\right)}.
	\label{eq:energy_angle}
\end{equation}
The total cross section is roughly 53~mb for low-energy scattering, and drops as $\hbar\omega_{\textrm{i}}$ approaches and exceeds the electron rest mass. 

The relativistic calculation of the photon energy and flux as a function of laboratory angle is most straightforward using the four-vector notation, in which the energy and momentum of the photon $\vec{k}$ and the Lorentz transformation matrix $L$ are:

\begin{align}
	\vec{k} &= \frac{\epsilon}{c}\left[
	\begin{array}{c}
		1 \\
	    \sin\theta_{\textrm{L}}\cos\phi_{\textrm{L}} \\
	    \sin\theta_{\textrm{L}}\sin\phi_{\textrm{L}} \\
	    -\cos\theta_{\textrm{L}}
	\end{array}
	\right] \\
	L\left(\vec{\beta}\right) &= \left[
	\begin{array}{cccc}
		\gamma & -\gamma \beta_x & \gamma \beta_y & \gamma \beta_z \\
		-\gamma \beta_x & 1 + \left(\gamma-1\right) \frac{\beta_x^2}{\beta^2} & \left(\gamma-1\right) \frac{\beta_x\beta_y}{\beta^2} & \left(\gamma-1\right) \frac{\beta_x\beta_z}{\beta^2} \\
		-\gamma \beta_y & \left(\gamma-1\right) \frac{\beta_y\beta_x}{\beta^2} & 1 + \left(\gamma-1\right) \frac{\beta_y^2}{\beta^2} & \left(\gamma-1\right) \frac{\beta_y\beta_z}{\beta^2} \\
		-\gamma \beta_z & \left(\gamma-1\right) \frac{\beta_z\beta_x}{\beta^2} & \left(\gamma-1\right) \frac{\beta_y\beta_z}{\beta^2} & 1 + \left(\gamma-1\right) \frac{\beta_z^2}{\beta^2} 
	\end{array}
	\right]
	\label{eq:fourvectors}
\end{align}
\noindent for photon energy $\epsilon = \hbar\omega$ and incident laser direction $(\theta_{\textrm{L}}, \phi_{\textrm{L}})$.  

To calculate the properties of the scattered photons, the following procedure is performed.
Initial laser photons $\vec{k}_{\textrm{i}}$ are boosted into the electron rest frame by applying the Lorentz transformation: $\vec{k}_{\textrm{i}}' = L(\vec{\beta}_0)\vec{k}_{\textrm{i}}$. 
(Primes indicate boosted quantities.)
The scattering is calculated using Eqs.~(\ref{eq:KleinNishina}) and (\ref{eq:energy_angle}), resulting in a new energy and trajectory for the photon, $\vec{k}_{\textrm{f}}'$.
The scattered photons are boosted back into the laboratory frame: $\vec{k}_{\textrm{f}} = L(-\vec{\beta}_0)\vec{k}_{\textrm{f}}'$.

Without loss of generality we choose $\phi_{\textrm{L}} = \pi/2$ and initial electron velocity $\vec{\beta}_0 = \beta_z \hat{z}$.
This results in a boosted photon with four-momentum

\begin{equation}
	\vec{k}_{\textrm{i}}' = \frac{\epsilon_{\textrm{i}}}{c}\left[
	\begin{array}{c}
		\gamma\left(1 + \beta\cos\theta_{\textrm{L}}\right) \\
	    0 \\
		\sin\theta_{\textrm{L}}\\
		-\gamma\left(\beta + \cos\theta_{\textrm{L}}\right) 
	\end{array}
	\right] \equiv 
	\frac{\epsilon_{\textrm{i}}'}{c}\left[
	\begin{array}{c}
		1 \\
		0 \\
		\sin\theta_{\textrm{L}}'\\
		-\cos\theta_{\textrm{L}}'
	\end{array}
	\right],
	\label{eq:ki_prime}
\end{equation}
\noindent where we have defined the boosted energy $\epsilon_{\textrm{i}}' = \epsilon_{\textrm{i}}\gamma\left(1 + \beta\cos\theta_{\textrm{L}}\right)$ and the boosted incident laser angle $\sin\theta_{\textrm{L}}' = \sin\theta_{\textrm{L}}/\gamma\left(1 + \beta\cos\theta_{\textrm{L}}\right)$.
Assuming $\theta_{\textrm{L}} \ll 1$ and $\gamma \gg 1$, we can neglect the off-axis contribution of the incident photon direction and approximate $\theta_{\textrm{L}}' \rightarrow 0$.
The resulting scattered photon has an energy $\epsilon_{\textrm{f}}' = \left(\omega_{\textrm{f}}'/\omega_{\textrm{i}}'\right)\epsilon_{\textrm{i}}'$ determined by Eq.~(\ref{eq:energy_angle}), with the scattered vector:

\begin{align}
	%\epsilon_{\textrm{f}}' &= \frac{\epsilon_{\textrm{i}} \gamma\left(1 + \beta\cos\theta_{\textrm{L}}\right)}{1+\frac{\epsilon_{\textrm{i}}}{m_{\textrm{e}}c^2}\gamma\left(1 + \beta\cos\theta_{\textrm{L}}\right)\left(1-\cos\theta_{\textrm{s}}\right)} \\
	\vec{k}_{\textrm{f}}' &= 
	\frac{\epsilon_{\textrm{f}}'}{c}
	\left[
	\begin{array}{c}
		1 \\
		\sin\theta_{\textrm{s}}\cos\phi_{\textrm{s}} \\
		\sin\theta_{\textrm{s}}\sin\phi_{\textrm{s}} \\
		-\cos\theta_{\textrm{s}}
	\end{array}
	\right]	.
	\label{eq:kf_prime}
\end{align}
Finally, transforming this back into the laboratory frame results in a final photon direction $\cos\theta$ and energy $\epsilon_{\textrm{f}}$,

\begin{align} 
	\cos\theta &= \frac{\beta - \cos\theta_{\textrm{s}}}{1-\beta\cos\theta_{\textrm{s}}} 	\label{eq:theta_final} \\
	\epsilon_{\textrm{f}} &= \frac{\epsilon_{\textrm{i}} \gamma^2\left(1 + \beta\right)\left(1 + \beta\cos\theta_{\textrm{L}}\right)}{1 + \gamma^2\beta\left(1+\beta\right)\left(1 - \cos\theta\right)+\frac{\epsilon_{\textrm{i}}}{m_{\textrm{e}}c^2}\gamma\left(1 + \beta\cos\theta_{\textrm{L}}\right)\left(1+\cos\theta\right)} .
	\label{eq:e_final}
\end{align}

 \begin{figure}
	\includegraphics[width=0.6\textwidth]{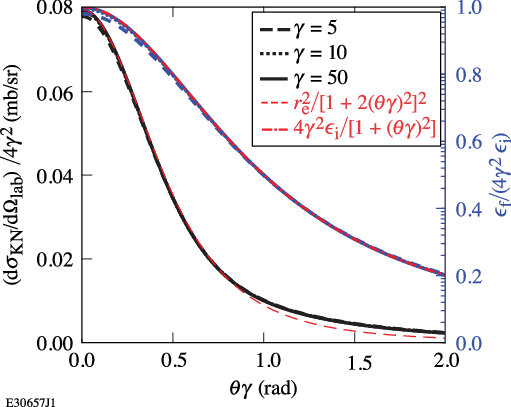} %{dsigma_E_vs_labangle.png}
	\caption{Scattering cross section normalized to $4\gamma^2$ (left, black) and scattered photon energy (blue, right) as functions of the scaled lab scattering angle $\theta\gamma$.\label{fig:frameshifted}}
\end{figure}

The variation in photon energy with angle arises from the second term in the denominator.  
(For optical photons, the third term is negligible for $\gamma \lesssim 10^5$.)
The median photon angle (produced at $\cos\theta_{\textrm{s}} \approx 0$) is $\theta = \cos^{-1}\beta$, or $\theta \approx 1/\gamma$ for $\gamma \gg 1$.
In this limit the second term is to lowest order $\gamma^2\theta^2$, as in Eq.~(\ref{eq:frequency}), and the energy of the median scattered photon is half of the maximum energy.
It can be shown from Eq.~(\ref{eq:theta_final}) that in the high-energy limit, the scattering angle $\cos\theta_{\textrm{s}} \xrightarrow[\gamma \gg 1]{} (\theta^2\gamma^2 - 1)/(\theta^2\gamma^2 + 1)$, which is a function only of $(\theta\gamma)^2$.
This explains the scattered photon energy and probability density of scattering scaling with this product, as shown in Fig.~(\ref{fig:analyticmodel}).

The Klein--Nishina cross section can be rewritten in the laboratory frame as $\textrm{d}\sigma_{\textrm{KN}}/\textrm{d}\Omega = (\textrm{d}\sigma_{\textrm{KN}}/\textrm{d}\Omega')(\textrm{d}\Omega'/\textrm{d}\Omega)$, with the Jacobian term $(\textrm{d}\Omega'/\textrm{d}\Omega) = \textrm{d}\cos\theta_{\textrm{s}}/\textrm{d}\cos\theta = \gamma^{-2}(1-\beta\cos\theta)^{-2}$.
An analytical form is straightforward to calculate from Eqs.~(\ref{eq:KleinNishina}) and (\ref{eq:theta_final}), and is plotted as a function of $\theta\gamma$ in Fig.~(\ref{fig:frameshifted}).
This is then integrated over laboratory solid angle to infer the photon fraction within a given acceptance angle ($f$) shown in Fig.~\ref{fig:analyticmodel}(b).

\subsection{Electron--Laser Intersection Volume}
\label{app:derivation:geometry}
\begin{figure}
	\includegraphics[width=\textwidth]{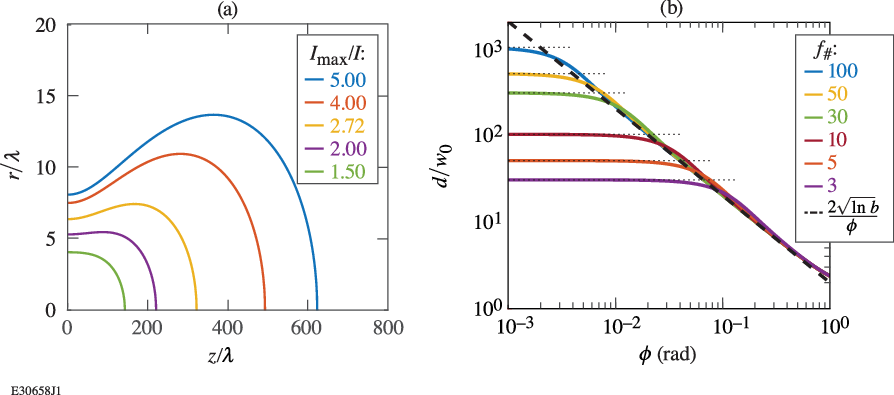} %{Fig_e_l_pathlength.png}
	\caption{(a) Intensity boundaries for a Gaussian laser pulse with $f_\# = 10~(w_0 = 6.37\lambda)$.  (b) Path length of an electron in the high-intensity region ($I_{\textrm{max}}/I = e$) for different focal numbers $f_\# = \pi w_0/2 \lambda$.  Limiting path lengths for on-axis scattering are $2 z_{\textrm{R}} \sqrt{e^2-1}$. To take advantage of long focal lengths ($f_\# > 20$), the impact angle must be close to 0. \label{fig:pathlength}}
\end{figure}

To maximize the number of scattering events at a given intensity, the electron path inside the focused laser spot should be as long as possible.
Assuming a Gaussian beam with radius at best focus $w_0 = 2 f_\#\lambda/\pi$ and Rayleigh length $z_{\textrm{R}} = \pi w_0^2/\lambda$, the radius of the beam is $w(z) = w_0 \sqrt{1 + (z/z_{\textrm{R}})^2}$.
The region with high intensity $(I > I_{\textrm{max}}/b)$ is then a volume with the boundary
\begin{equation}
	\left(\frac{r}{w_0}\right)^2 = \left(1 + \frac{z^2}{z_{\textrm{R}}^2}\right)\left[\ln{b} - \frac{1}{2}\ln\left(1+\frac{z^2}{z_{\textrm{R}}^2}\right)\right],
	\label{eq:highI_bdry}
\end{equation}
\noindent as shown in Fig.~\ref{fig:pathlength}(a).
The maximum path length for an electron transiting this boundary depends on the angle of incidence $\phi$ [from Eq.~(\ref{eq:frequency})].
Taking $r = z\tan\phi$, we can solve for the path length $d = 2z/\cos\phi$ as shown in Fig.~\ref{fig:pathlength}(b).
The length of the interaction grows with reduced impact angle as $d \approx 2 \sqrt{\ln b}/\phi$, up to a limiting value that depends on the Rayleigh length as $d_{\textrm{max}} = 2 z_{\textrm{R}} \sqrt{b^2-1}$.  
[The scaling in Eq.~(\ref{eq:N_source}) uses an intensity boundary of $b = \sqrt{2}$, such that $d_{\textrm{max}} = 2 z_{\textrm{R}}$.]
To take advantage of these long interaction volumes, however, requires a collision angle close to $\phi = 0$.
For example, at $\phi = 1^\circ~(0.017~\textrm{rad})$, there is no additional increase in the normalized path length with $f_\# \gtrsim 16$, which corresponds to  $w_0/\lambda > 10$.
This conclusion depends on the use of Gaussian beams: more-realistic focusing schemes (for example, flat profiles in the far field) must be evaluated in future designs.

If a flying-focus pulse is used, the length of the intense region that co-moves with the electron packet is approximately twice the Rayleigh length, and the portion of the electron packet that can scatter at high intensity is limited to this length.
%%% NEEDS UPDATE:  w_0 = (2 f lambda / pi) for diffraction limited!  Not (f lambda / 2)!
For diffraction-limited focusing, $w_0 = 2 f_\# \lambda/\pi$, the electron packet width is limited to $\tau \leq 4 f_\#^2\lambda/\pi c = (f_\#^2) 0.00447$~ps.
For reasonable values of $f_\#$, this is much shorter than the values described in Sec.~\ref{sec:sub:electrons}, and severely limits the charge contained in the packets.
In this case the photoinjector would be optimized for low emittance and the laser will be designed to achieve long interaction lengths and high intensities on axis.

\end{document}